\documentclass[12pt]{article}
\usepackage[english]{babel}
\usepackage[T1]{fontenc}
\usepackage{latexsym}
\usepackage{amsfonts}
\usepackage{epsfig}
\usepackage{dsfont}
\usepackage{cprform}

\begin{document}
\def\op#1{\mathcal{#1}}
\def\bfnull{\relax{\rm O \kern-.635em 0}}
\def\dop{{\rm d}\hskip -1pt}
\def\a{\alpha}
\def\b{\beta}
\def\g{\gamma}
\def\d{\delta}
\def\e{\epsilon}
\def\ve{\varepsilon}
\def\t{\theta}
\def\l{\lambda}
\def\m{\mu}
\def\n{\nu}
\def\pg{\pi}
\def\r{\rho}
\def\s{\sigma}
\def\t{\tau}
\def\z{\zeta}
\def\c{\chi}
\def\p{\psi}
\def\o{\omega}
\def\G{\Gamma}
\def\D{\Delta}
\def\T{\Theta}
\def\L{\Lambda}
\def\Pg{\Pi}
\def\S{\Sigma}
\def\O{\Omega}
\def\pb{\bar{\psi}}
\def\cb{\bar{\chi}}
\def\lb{\bar{\lambda}}
\def\Pii{\mathcal{P}}
\def\Q{\mathcal{Q}}
\def\K{\mathcal{K}}
\def\A{\mathcal{A}}
\def\N{\mathcal{N}}
\def\F{\mathcal{F}}
\def\Gi{\mathcal{G}}
\def\Ci{\mathcal{C}}
\def\oL{\overline{L}}
\def\oM{\overline{M}}
\def\wk{\widetilde{K}}
\def\hb{\overline{h}}
\def\eq#1{(\ref{#1})}
\newcommand{\be}{\begin{equation}}
\newcommand{\ee}{\end{equation}}
\newcommand{\ba}{\begin{eqnarray}}
\newcommand{\ea}{\end{eqnarray}}
\newcommand{\ban}{\begin{eqnarray*}}
\newcommand{\ean}{\end{eqnarray*}}
\newcommand{\nn}{\nonumber}
\newcommand{\nin}{\noindent}
\newcommand{\fgl}{\mathfrak{gl}}
\newcommand{\fu}{\mathfrak{u}}
\newcommand{\fsl}{\mathfrak{sl}}
\newcommand{\fsp}{\mathfrak{sp}}
\newcommand{\fusp}{\mathfrak{usp}}
\newcommand{\fsu}{\mathfrak{su}}
\newcommand{\fp}{\mathfrak{p}}
\newcommand{\fso}{\mathfrak{so}}
\newcommand{\fg}{\mathfrak{g}}
\newcommand{\fr}{\mathfrak{r}}
\newcommand{\fe}{\mathfrak{e}}
\newcommand{\rE}{\mathrm{E}}
\newcommand{\rSp}{\mathrm{Sp}}
\newcommand{\rSO}{\mathrm{SO}}
\newcommand{\rSL}{\mathrm{SL}}
\newcommand{\rSU}{\mathrm{SU}}
\newcommand{\rUSp}{\mathrm{USp}}
\newcommand{\rU}{\mathrm{U}}
\newcommand{\rF}{\mathrm{F}}
\newcommand{\R}{\mathbb{R}}
\newcommand{\C}{\mathbb{C}}
\newcommand{\Z}{\mathbb{Z}}
\newcommand{\Hb}{\mathbb{H}}
\def\oL{\overline{L}}
\def\mW{\mathcal{W}}


\begin{titlepage}
\begin{center}

\rightline{\small IFT-UAM/CSIC-06-62}
\vskip 1cm

{\Large \bf On the underlying $E_{11}$ symmetry\\[2mm] of the $D=11$ Free Differential Algebra}
\vskip 1.2cm

{\bf 
Silvia Vaul\`a }
\vskip 0.2cm
{\it Instituto de F\'{\i}sica Te\'orica UAM/CSIC\\
Facultad de Ciencias C-XVI,  C.U.~Cantoblanco,  E-28049-Madrid, Spain}

\vskip 0.4cm

 {\tt silvia.vaula@uam.es} 
\vskip 0.4cm

\end{center}

\vskip 1cm

\begin{center} {\bf ABSTRACT }\end{center}
\vskip 0.4cm

We study the reduction of the Free Differential Algebra (FDA) of $D=11$ supergravity to an ordinary algebra. We show that in flat background and with vanishing three--form field strength, the corresponding minimal FDA can be reduced to an In\"on\"u--Wigner contraction of Sezgin's $M$--Algebra. We also prove that in flat background but with a non trivial three--form field strength, the bosonic FDA can be reduced to the lowest levels of $E_{11}$. This result suggests that the $E_{11}$ symmetries, which act on perturbative states as well, are already encoded in the $D=11$ FDA and are made explicit when the theory is formulated on a enlarged group manifold.

\noindent

\vfill

\end{titlepage}

\section{Introduction}

Eleven dimensional  supergravity \cite{Cremmer:1978km}  and its symmetries constitute  a privileged access to the study of $M$--Theory as the former is believed to be the low energy limit of the latter. There are some hints indicating the possibility that the symmetry group of $M$--Theory is represented by infinite dimensional Ka\v{c}--Moody group $E_{11}$. 

The first hint in this direction is that the $U$--duality group of  maximal supergravities in $D\geq3$ dimensions  is given by the exceptional group series\footnote{We use the notation $E_1=\mathbb{R}$, $E_2={\rm  GL}(2,\mathbb{R})$, $E_3={\rm  SL}(2,\mathbb{R})\times {\rm  SL}(3,\mathbb{R})$,  $E_4={\rm  SL}(5,\mathbb{R})$, $E_5={\rm O}(5,5)$. With $E_{p(p)}$ we indicate the real section of $E_{p}$ with $p$ non compact generators, that is with the maximum number of non compact generators. For simplicity of notation in the following  with will indicate it just with $E_{p}$.} $E_{11-D(11-D)} $\cite{Cremmer:1978ds}. Maximal supergravities in $D$ dimensions can be obtained form $T_{11-D}$ compactification of the $D=11$ supergravity, nevertheless the ${\rm SL}(11-D,\mathbb{R})$ symmetries induced by  the $T_{11-D}$ are just part of the exceptional groups symmetries they feature. \\
Going down to $D<3$ leads to $U$--duality groups which are no more finite dimensional \cite{Julia:1981wc} but they are the Ka\v{c}--Moody groups $E_{9(9)}$,  $E_{10(10)}$ and  $E_{11(11)}$,  that is the affine extension, the over extension and the very extension of $E_{8(8)}$ respectively.
This suggests that  a plausible scenario is that the $D=11$ supergravity itself features an $E_{11(11)}$ hidden symmetry.

There have been several proposals for  non linear formulation of the $D=11$ supergravity \cite{Cremmer:1998px,West:2000ga}, in particular as non--linear realizations of $E_{10}$ or  $E_{11}$ \cite{West:2001as,Damour:2005zs,deBuyl:2005mt,Damour:2006xu}. In \cite{Damour:2005zs,deBuyl:2005mt,Damour:2006xu} an action based on the coset space $E_{10}/K(E_{10})$ was considered,  $K(E_{10})$  being the maximally compact subgroup of  $E_{10}$. This is a natural generalization of the $U$--duality invariant action of the scalar $\s$--model  $E_{D-11(D-11)}/K(E_{D-11(D-11)})$ in $D$--dimensional supergravity.

Implications of this underlying larger symmetries can also be found in some cosmological solutions exhibiting billiard phenomenon \cite{Damour:2000wm,Damour:2002et} .

More recently, it has been shown \cite{Bergshoeff:2006qw} that the supersymmetry transformation laws of IIA supergravity, after suitable field dependent redefinitions of the gauge fields and the gauge parameters, become linear in the gauge fields while the resulting gauge algebra reproduces the lowest levels of $E_{11}$. 

In the present paper we will look for $E_{11}$ symmetries in  $D=11$ supergravity, taking as a starting point the $D=11$ Free Differential Algebra  (FDA) \cite{D'Auria:1982nx}.\\
Free Differential Algebras \cite{sullivan}, which are a generalization of the concept of Lie algebras, turn out to be relevant for the construction of higher dimensional supergravities  where the supermultiplets contain $p$--form potentials, with $p>1$. In fact, for $p=1$ the gauge potentials are associated to the one--forms dual to the Lie algebra generators of the symmetry group of the theory \cite{cubo}.  For $p>1$ the $p$--form potentials are associated to the $p$--form generators of a suitable Free Differential Algebra encoding the symmetries of the theory.\\
The minimal $D=11$ FDA consists of the Maurer--Cartan equations of the $D=11$ super Poincar\'e algebra plus a generalized Maurer--Cartan equation for the three--form $C$. It is well known  \cite{D'Auria:1982nx} that the minimal $D=11$ FDA can be reduced to a set of ordinary Maurer--Cartan equations describing an extension of the $D=11$ super Poincar\'e algebra, via the expansion of the three--form $C$ in terms of the one--forms dual to the generators. \\
Inspired by this observation we consider Sezgin's $M$--Algebra \cite{Sezgin:1996cj} which is the most general extension of the $D=11$ super Poincar\'e algebra and check whether the minimal $D=11$ FDA can be reduced to it. As we are considering the minimal $D=11$ FDA, that is when all the curvatures and field strengths are zero\footnote{For a definition of minimal and contractible FDA and their relations with $D=11$ supergravity, see e.g.\cite{Fre:1984pc}. In the present paper we will have a minimal FDA whenever the curvatures and field strengths, representing the contractible generators, are vanishing.}, we have to be consistent with a flat background, and therefore forced to avoid non--commuting translations. This constraint leads to an In\"on\"u--Wigner contraction of the $M$--Algebra which we will show to be a reduction of the minimal $D=11$ FDA. This is the main result we will present: beside showing how the $M$--Algebra naturally arises in $D=11$ supergravity, we provide the general structure for a candidate flat background superfivebrane Wess--Zumino term.

The next step is  the analysis of the complete $D=11$ FDA, that is in the presence of nonzero curvatures and field strengths (= contractible generators). For this more general case it has not yet been proven whether exists the possibility to reduce it to an algebra, neither we will prove it now. We will limit ourselves to consider the simplest case, that is flat background where just the super filed strength $F$ of the three--form is present, and to analyze it at the bosonic level.\\
In spite of the simplification, this example indicates that in order to find an expansion for $F$ we have to consider the automorphism algebra acting on the $M$--Algebra and expand $F$ in terms of its dual generators.  The automorphism algebra  \cite{Barwald:1999is,West:2004iz} needed for the bosonic part of  $F$ in flat background, turns out to coincide with the lowest levels of $E_{11}$. This partial result suggests that  the  $E_{11}$ symmetry is already encoded in the $D=11$ FDA and in order to make it manifest one needs to reduce the FDA to an algebra. 
  
The paper is organized as follows:

In section 2, after recalling the basic ideas about minimal Free Differential Algebras, we perform the reduction of the minimal $D=11$ FDA to the maximal In\"on\"u--Wigner contraction of the $M$--Algebra allowing for commuting translations.

In section 3 we present the general  $D=11$ FDA  and discuss the treatment of the contractible generators. We focus on the bosonic components of the three--form super field strength $F$ and show that the bosonic  $D=11$ FDA in flat background can be reduced to the lowest levels of $E_{11}$.

In section 4 we draw our conclusions and discuss future perspectives.

In appendix A we list the Fierz identities needed for the reduction of the minimal $D=11$ FDA.

In appendix B we write explicitly the system of equations for the coefficients of the expansion of the three--form $C$ and its solutions.

In appendix C we report the rheonomic parametrizations for the $D=11$ supercurvatures and their relation with supersymmetry transformation laws.

\section{The composite nature of the $D=11$ three--form}

\subsection{Reduction of the minimal $D=11$ FDA to an ordinary algebra}\label{riccardopietro}

It is well known that there are two equivalent ways to locally characterize a (super) Lie group manifold $\mathcal{G}$.\\
One is by means of  the commutation relation between the basis elements $\{T_A\}$ of its tangent space  $\mathcal{T}\mathcal{G}$,  $A=1,\dots dim\,\mathcal{G}$
\be [T_A,\,T_B\}=C^C_{AB}T_C\,,\ee
that is its associated (super) Lie algebra, or via the Maurer--Cartan equations (MCE) on the basis elements  $\{\m^A\}$  of its cotangent space  $\mathcal{T}^{*}\mathcal{G}$ 
\be d\m^A-\frac12C^A_{CB}\,\m^B\wedge\m^C=0\,,\label{MCE}\ee
where the dual basis is defined canonically by $\m^A(T_B)=\d^A_B$.\\
The Jacobi identities 
\be C^A_{\ B[C}C^B_{\ DE\}}=0\label{jaco0}\ee
are respectively obtained from
\be [T_{[A},\,[T_B,\,T_{C\}}\}\}=0;\quad\quad d\left(d\m^A-\frac12C^A_{CB}\,\m^B\wedge\m^C=0\right)\label{jaco}\ee   
The formulation in terms of Maurer--Cartan equations turns out to be more appropriate for the construction of a supergravity theory with the geometric approach \cite{cubo}, since the fundamental fields are directly associated to the one--forms $\m^A$.\\
In the specific case of $D=11$ supergravity  \cite{Cremmer:1978km}, on--shell supersymmetry is realized on the following set of fields 
\be(g_{\m\n},\,\p_\m,\,C_{\m\n\r})\quad\quad \m,\,\n,\,\r=0,\dots 10\, .\ee 
One can describe the gravitational degrees of freedom $g_{\m\n}$ by means of the vielbein $V^a$ and the spin connection $\o^{ab}$  ($a,\,b=0,\dots10$), defined as the one--forms dual to the translation generators $P_a$ and the Lorentz generators $M_{ab}$ respectively; the gravitino $\psi$ is defined as the one--form dual to the supersymmetry generator $Q$. In the Minkowski vacuum they satisfy the Maurer--Cartan equations of the  eleven dimensional super Poincar\'e algebra\footnote{Here and in the following the symbol of wedge product "$\wedge$" between $p$--forms is suppressed in order to simplify the notation.} \cite{D'Auria:1982nx} whose closure under $d$ differentiation, \eq{jaco},  is trivially checked:
 \ba
&&T^a\equiv DV^a-\frac i2\pb\g^a\p=0\label{DV}\\
&&\mathcal{R}^{ab}\equiv d\o^{ab}-\o^a_{~c}\o^{cb}=0\label{do}\\
&&\r\equiv D\p=0\label{Dp}\,,
\ea
Equations \eq{DV} and \eq{do} respectively define the supertorsion $T^a$ and the riemaniann supercurvature $R^{ab}$ of the superspace, while equation \eq{Dp} defines the gravitino supercurvature $\r$.  The structure of Maurer--Cartan equations \eq{MCE} of \eq{DV}--\eq{Dp} is due to the fact that in the Minkowski background $T^a=\mathcal{R}^{ab}=\r=0$. The  "covariant derivatives" are defined as follows:
\be DV^a\equiv dV^a-\o^{ab}V_b;\quad\quad D\p\equiv d\p-\frac14\o^{ab}\g_{ab}\p\label{coder}\ee
Let us point out that, at present, \eq{coder} is just a formal definition which does not have the meaning of  covariant derivative. Indeed,  the one--forms $(V^a,\,\o^{ab},\,\p)$ are all defined as sections of the cotangent space $\mathcal{T}^{*}\mathcal{G}$ and depend on the coordinates $(x^a,\,x^{ab},\,\theta)$ of the supergroup manifold. More precisely,  $x^a$ are the ordinary space--time coordinates associated to translations, while $\theta$ is a 32 component Majorana spinor describing the grassmannian coordinates which together with the $x^a$ parametrize the ordinary $D=11$ superspace; the $x^{ab}$ are associated to the Lorentz subgroup and  physical fields should not depend on them. In order to recover the ordinary superspace, where $\o^{ab}$ plays the role of spin connection and the fields only depend  on the superspace coordinates $\{x^a,\,\theta\}$, one has to impose horizontality with respect to the subgroup $\rm{SO}(1,10)$ \cite{cubo}; this is tantamount  saying that the theory is formulated on a coset manifold $\mathcal{G}/\mathcal{H}$ where $\mathcal{G}=$ super--Poincar\'e and $\mathcal{H}=$Lorentz.

Here and in the following, we will always refer to the group manifold $\mathcal{G}$  without imposing horizontality with respect to Lorentz , which is meant to be imposed in a second step if one wants to recover a superspace formulation; nevertheless we will refer to  "covariant derivatives", "torsion", "curvature", etc. keeping in mind the previous disclaimer. This applies as well when we will consider extensions $\tilde\mathcal{G}$ of  the group manifold $\mathcal{G}$; horizontality with respect to its maximal compact subgroup $\tilde\mathcal{H}\equiv K(\tilde\mathcal{G})$ is meant to be imposed afterwards.

In order to be able to include the three--form field $C_{\m\n\r}$ one needs to extend the super Lie  algebra \eq{DV}--\eq{Dp} to a Free Differential Algebra \cite{D'Auria:1982nx}.  For our purposes we need an $\rm {SO}(1,10)$ singlet four--form that is closed but not exact in $\L_4(\mathcal{T}^*\mathcal{G})$.  The only choice compatible with a linear realization of supersymmetry is \cite{cubo}:
\be w^{(4)}=\pb\g_{ab}\p V^aV^b\ee
whose closure $dw^{(4)}=0$ can be checked using the Fierz identity \eq{4lin1}, and equations \eq{DV} and \eq{Dp}. 

The four--form $w^{(4)}$ is not exact on  $\L_4(\mathcal{T}^*\mathcal{G})$, but by enlarging the group manifold $\mathcal{G}$ to a suitable manifold $\tilde\mathcal{G}$, one can  introduce a three--form $C$ on $\L_3(\mathcal{T}^*\tilde\mathcal{G})$ which satisfies:
\be F\equiv dC-\frac 12 \pb\g_{ab}\p V^aV^b=0\label{dC}\ee
such that \eq{DV}--\eq{Dp} together with \eq{dC} are closed under $d$ differentiation: the resulting structure is a minimal FDA.

One can further extend the FDA \eq{DV}--\eq{Dp}, \eq{dC} in order to include the six--form dual to $C$ (see \cite{D'Auria:1982nx,Fre:1984pc} for a complete discussion), but we will not consider such a formulation, as its treatment is beyond the scope of the present paper.

Since in the minimal FDA the supercurvatures of the $D=11$ fields are identically zero $T^a=\mathcal{R}^{ab}=\r=F=0$, and in particular $\mathcal{R}^{ab}=0$,  we can set the trivial spin connection to zero $\o^{ab}=0$, thus reducing  the covariant derivatives \eq{coder}  to ordinary ones. Furthermore we can also neglect equation \eq{do} and  reduce to the following minimal FDA, based on an extension $\tilde\mathcal{G}_0$ of the super translational group $\mathcal{G}_0$ manifold (equations  \eq{dVf}, \eq{dpf} below)
\ba
&&dV^a-\frac i2\pb\g^a\p=0\label{dVf}\\
&&d\p=0\label{dpf}\\
&&dC-\frac 12 \pb\g_{ab}\p V^aV^b=0\label{dCf}
\ea
It is quite natural to wonder if $\tilde\mathcal{G}_0$ can be realized as a group manifold. This is tantamount to say that, denoting by
\be\{\s^i\} \supset \{V^a\,,\p\},\quad i=1\dots  dim\, \tilde\mathcal{G}_0\ee
the basis of $\mathcal{T}^*\tilde\mathcal{G}_0$, they satisfy
\be d\s^i-\frac12 c^i_{\ kj}\,\s^j\wedge\s^k=0,\quad\quad d\left(d\s^i-\frac12 c^i_{\ kj}\,\s^j\wedge\s^k\right)=0\label{newG}\ee
with constant $c^i_{\ kj}$, in such a way that \eq{dVf}, \eq{dpf} are included in \eq{newG}.\\
If this is the case it is possible to express $C$ as 
 \be C=K_{ijk}\s^i\s^j\s^k\label{Csplit}\, ,\ee
where the constants $K_{ijk}$ are determined by imposing \eq{dCf} and using \eq{newG}.\\ 
This corresponds to the statement  that  \eq{newG} is an algebra equivalent to the FDA \eq{dVf}--\eq{dCf}  \cite{D'Auria:1982nx}. The group $\tilde\mathcal{G}_0$ corresponding to the FDA \eq{dVf}--\eq{dCf} may be not unique \cite{deAzcarraga:2005jd}.

In fact, the question whether the minimal $D=11$ FDA is equivalent to an ordinary algebra, was first addressed in \cite{D'Auria:1982nx} and solved introducing as new one--forms two bosons $B^{ab}$, $B^{a_1\dots a_5}$ and one fermion $\eta$. 
The two new bosonic one--forms turn out to be dual to the central charges generators $Z_{ab}$, $Z_{a_1\dots a_5}$ 
\be \{Q_\a, Q_\b\}= i\g^a_{\a\b}P_a+\g^{ab}_{\a\b}Z_{ab}+i\g^{a_1\dots a_5}_{\a\b}Z_{a_1\dots a_5}\label{central}\ee
Afterwards \cite{Bandos:2004xw,Bandos:2004ym} it was shown that in the framework of \cite{D'Auria:1982nx} there exist  a whole class of solutions, in particular in the absence of $B^{a_1\dots a_5}$. More recently \cite{Castellani:1995gz,Castellani:2005vt,Castellani:2006jg}  a different approach was proposed that make use of ``extended'' Lie derivatives  along antisymmetric tensors. A solution was derived  in terms of $B^{ab}$ and a fermion $\eta^a$ carrying a Lorentz index plus a further bosonic generator\footnote{$Z^{\a\b}$ in the notations of \cite{Castellani:2005vt}.} $\S_{\a\b}$. It was also pointed out that this latter  $\S_{\a\b}$ generator vanishes for null curvatures, in particular for a trivial spin connection.

In spite of the evidence that there is no unique algebra which corresponds to the minimal $D=11$ FDA, one can ask which is the biggest extension $\mathcal{T}\tilde\mathcal{G}_0$ of the super translational algebra \eq{dVf}, \eq{dpf} which can satisfy \eq{dCf}. The most general extension of the super translational algebra is represented by the $M$--Algebra proposed by Sezgin \cite{Sezgin:1996cj}.  Encouraged by the fact that all the previous results  \cite{D'Auria:1982nx,Castellani:2005vt,Bergshoeff:1995hm,Sezgin:1995bh,Chryssomalakos:1999xd} represent particular cases of the  $M$--Algebra, in the next subsection \ref{M} we will investigate if the $D=11$ FDA can be reduced to the latter. 


\subsection{The $M$--Algebra from the $D=11$ minimal FDA}\label{M}
In this section we are going to discuss whether the $D=11$ minimal FDA can be reduced to the $M$--Algebra  \cite{Sezgin:1996cj}. In order to do that, we have to write \eq{Csplit} in terms of the one--forms dual to the $M$--Algebra's generators and afterwards see if there exist constants $K_{ijk}$ such that \eq{dCf} is satisfied once the Maurer--Cartan equation of the $M$--Algebra are used.
 
 As is evident from equation \eq{dCf}, the expression for the three--form $C$ can be just determined up to closed three--forms. Therefore we will not include in our ansatz for $C$ forms like \cite{Sezgin:1996cj}
\be w^{(3)}=V^a\,\pb\g_a\p-(1-\l-\t)\,B^a\,\pb\g_a\p+i\frac{\l}{10}B^{ab}\pb\g_{ab}\p+\frac{\t}{720}B^{a_1\dots a_5}\pb\g_{a_1\dots a_5}\p\label{chiusa}\ee
whose closure can be checked using the Fierz identity \eq{fierz1}

The $M$--Algebra \cite{Sezgin:1996cj} can be obtained by subsequent central extensions of the algebra \eq{dVf}, \eq{dpf}; in terms of commutators this also reads
\ba
\{Q_\a, Q_\b\} &=& i\g^a_{\a\b}\ P_a\label{QQ}\\
\left[Q_\a,P_b\right ]&=&0\label{QP}\\
\left[P_a,P_b\right]&=&0\label{PP}
\ea  
The first extension introduces the generators $Z_a$, $Z_{ab}$, $Z_{a_1\dots a_5}$ on the r.h.s of \eq{QQ}, thus giving \eq{QQ+} below.
The second extension introduces the generators $\S_\a$, $\S^{a}_\a$, $\S^{a_1\dots a_4}_{\a}$  on the r.h.s of \eq{QP}, thus giving \eq{QP+}, but as well non trivial commutators between the $Q_\a$ and the generators introduced in the previous step (which cease to be central) \eq{QZ1}--\eq{QZ5}:    
\ba
\{Q_\a, Q_\b\}\! \!&=&\!\! i \g^a_{\a\b}\ P_a + i\g^{a}_{\a\b}\ Z_a+\g^{ab}_{\a\b}\ Z_{ab}+i\g^{a_1\dots a_5}_{\a\b}\ Z_{a_1\dots a_5}\label{QQ+} \\
\left[Q_\a, P_a\right]\!\! &=&\!\! i\g_{a\a\b}\ \S^\b +\g_{ab\a\b}\ \S^{b\b}+\g_{ab_1\cdots b_4\a\b}\ \S^{b_1\cdots b_4\b}\label{QP+} \\
\left[Q_\a, Z^a\right]\!\! &=&\!\! -i\,(1-\l-\tau)~\g^a_{\a\b}\ \S^\b\label{QZ1}\\
\left[Q_\a, Z^{ab}\right]\!\! &=&\!\! \frac{\l}{10} \g^{ab}_{\a\b}\ \S^\b +i\g^{a}_{\a\b}\ \S^{b\b}-6i\g_{\a\b}^{cd}\S^{abcd\b}\label{QZ2}\\ 
\left[Q_\a, Z^{a_1\dots a_5}\right]\!\! &=&\!\! i\frac{\tau}{720} \g^{a_1\dots a_5}_{\a\b}\ \S^\b +\g^{a_5}_{\a\b}\ \S^{a_1\dots a_4\b}\label{QZ5}\\
{\rm zero\ otherwise.}&&\nn
\ea
The next extension would introduce new generators $\S_{\a\b}$, $\S_{\a\b}^{\a_1\dots a_3}$ on the r.h.s. of \eq{PP} and also further non--trivial commutators \cite{Sezgin:1996cj}.

As we are considering a flat background we cannot consider non commuting translations; furthermore it was pointed out before \cite{Castellani:2005vt}  that  $\S_{\a\b}$ is proportional to $\o^{ab}$ and therefore can be set to zero in a flat background. As a consequence we will not consider this further extension and take the algebra \eq{QQ+}--\eq{QZ5} as the candidate extension $\mathcal{T}\tilde\mathcal{G}_0$ for the reduction of the minimal FDA \eq{dVf}--\eq{dCf} (note that some factors in \eq{QQ+}--\eq{QZ5} change with respect to \cite{Sezgin:1996cj} as we are using the conventions of \cite{D'Auria:1982nx}).\\
Introducing the canonical basis on $\mathcal{T}^*\tilde\mathcal{G}_0$
\ba B^{a_1\dots a_p}Z_{b_1\dots b_p}&=&\d^{a_1\dots a_p}_{b_1\dots b_p}\\
\eta^\a_{a_1\dots a_{p-1}}\S_\b^{b_1\dots b_{p-1}}&=&\d^\b_\a\, \d^{b_1\dots b_{p-1}}_{a_1\dots a_{p-1}};\quad\quad\quad p=1,\,2,\,5
\ea
and suppressing the spinorial index $\a$, we can write the following Maurer--Cartan equations:
\ba
&&d V^a-\frac i2 \pb\g^a\p=0\label{MC1}\\
&&d B^a-\frac i2 \pb\g^a\p=0\label{MC2}\\
&&d B^{a_1a_2}-\frac 12\pb\g^{a_1a_2}\p=0\label{MC3}\\
&&d B^{a_1\dots a_5}-\frac i2\pb\g^{a_1\dots a_5}\p=0\label{MC4}\\
&&d\p=0\label{MC5}\\
&&d\eta+i\d_0\g_a\p V^a\!+\!i\d_3\g_a\p B^a+\d_1\g_{ab}\p B^{ab}\!+i\d_2\g_{a_1\dots a_5}\p B^{a_1\dots a_5}=0\label{MC6}\\    
&&d\eta^{a_1}+\g_1\g^{a_1a_2}\p V_{a_2}+\g_2\g_{a_2}\p B^{a_1a_2}=0\label{MC7}\\
&&d\eta^{a_1\dots a_4}+\pi_1\g^{a_1\dots a_5}\p V_{a_5}+\pi_2\g_{a_5}\p B^{a_1\dots a_5}+\pi_3 \g^{[a_1a_2}\p  B^{a_3a_4]}=0\label{MC8}\ea
In order to check the closure \eq{jaco} of \eq{MC1}--\eq{MC8} we need the Fierz identities \eq{fierz1}-- \eq{fierz3} which impose the following relations:

\ba
&&\d_0+\d_3+10\d_1-720\d_2=0\label{cl1}\\
&&\g_2=i\g_1\label{cl2}\\
&&\pi_1=\pi_2\label{cl3}\\
&&6i\pi_2+\pi_3=0\label{cl4}
\ea
There is some redundancy in the parameters introduced in \eq{MC1}--\eq{MC8}, since e.g. $\d_0$, $\g_1$ and $\pi_1$ can be reabsorbed in the definitions of
$\eta$, $\eta^a$ and $\eta^{a_1\dots a_4}$ respectively.  Due to this redundancy we are able to chose the same normalization as in  \cite{Sezgin:1996cj}, i.e.:
\ba
&& \d_0=-\frac 12;\quad \d_1=\frac{\l}{20};\quad \d_2=-\frac{\t}{1440};\quad \d_3=\frac 12 (1-\l-\t);\label{n1}\\
&&\g_1=-\frac 12;\quad \g_2=-\frac i2;\label{n2}\\
&&\pi_1=\frac 12;\quad \pi_2=\frac 12;\quad \pi_3=-3i \label{n3}
\ea

Let us then consider the following expansion for the three--form $C$ in terms of the set of one forms $\{V^a,\, B^{a_1\dots a_p},\, \p,\,\eta^{a_1\dots a_{p-1}} \}$, $ p=1,\,2,\,5$: 
\ba
C^{(3)}\!\!&=&\!\!\a_0 B^{a_1a_2} V_{a_1} V_{a_2}+
\a_1 B^{a_1}_{~a_2} B^{a_2}_{~a_3} B^{a_3}_{~a_1}+
\a_2B_{b_1a_1\dots a_4} B^{b_1}_{~b_2}B^{b_2a_1\dots a_4}+\nn\\[1mm]
&&+\a_3\epsilon_{a_1\dots a_5 b_1\dots b_5 m}B^{a_1\dots a_5}B^{b_1\dots b_5}V^m+\nn\\[1mm]
&&+\a_4\epsilon_{a_1\dots a_6 b_1\dots b_5 m}B^{a_1a_2a_3p_1p_2}B^{a_4a_5a_6}_{p_1p_2}B^{b_1\dots b_5}+\nn\\[1mm]
&&+\a_5\epsilon_{a_1\dots a_5 b_1\dots b_5 m}B^{a_1\dots a_5}B^{b_1\dots b_5}B^m+\nn\\[1mm]
&&+\a_6B^{ab}B_aV_b+\a_7B^{ab}B_aB_b+\nn\\[1mm]
&&+\b_1\pb\g_{a_1a_2}\eta^{a_1} V^{a_2}+
\b_2\pb\g^{a_1} \eta^{a_2}B_{a_1a_2}+
\b_3\pb\g_{a1}\eta_{a_2\dots a_5} B^{a_1\dots a_5}+\nn\\[1mm]
&&+\b_4\pb \g_{a_1a_2}\eta^{a_1a_2a_3a_4} B_{a_3a_4}+
\b_5\pb \g_{a_1\dots a_5}\eta^{a_1\dots a_4}V^{a_5}+\b_6\pb \g_{a_1\dots a_4}\eta_{a_5} B^{a_1\dots a_5}+\nn\\[1mm]
&&+\b_7\pb \g_{a_1a_2a_3}\eta_{a_3} B^{a_1a_2}+\b_8\pb\g_{a_1a_2}\eta^{a_1} B^{a_2}+\b_9\pb \g_{a_1\dots a_5}\eta^{a_1\dots a_4}B^{a_5}+\nn\\[1mm]
&&+i\hat\beta_1\pb\g_a\eta V^a+\hat\beta_2\pb\g_{ab}\eta B^{ab}+i\hat\beta_3\pb\g_{a_1\dots a_5}\eta B^{a_1\dots a_5}+i\hat\beta_4\pb\g_a\eta B^a\label{Cparam}
\ea
Imposing \eq{dCf} on \eq{Cparam}, using the MCE \eq{MC1}--\eq{MC8} and the Fierz identities \eq{fierz1}--\eq{fierz3} one obtained a system \eq{ppVV}--\eq{sistema} of second order equations for the coefficients $K_{ijk}$ in the expansion \eq{Cparam} that we report in appendix \ref{sol}.\\
In the following we discuss its solutions for different values of $p$; the coefficients in \eq{Cparam} for each case are listed in appendix \ref{sol}.
\begin{itemize}
\item[]$\mathbf{p=1}$: there are no solutions to \eq{dCf}.

\item[]$\mathbf{p=2}$: we retrieve the algebra of \cite{Castellani:2005vt} which is given\footnote{The normalization chosen in \cite{Castellani:2005vt} is $\g_1=2$.} by \eq{MC1}, \eq{MC3}, \eq{MC5}, \eq{MC7} .

\item[]$\mathbf{p=5}$: due to the Fierz identity \eq{fierz3} it is impossible to satisfy the closure of \eq{MC8} without introducing the generators $p=2$.

\item[]$\mathbf{p=1,\,2}$: the algebra is given by \eq{MC1}, \eq{MC2}, \eq{MC5}, \eq{MC6}. 

\item[]$\mathbf{p=1,\,5}$: once again, due to the Fierz identity \eq{fierz3} it is impossible to satisfy the closure of \eq{MC6} without introducing the generators $p=2$.

\item[]$\mathbf{p=2,\,5}$: the algebra is given by \eq{MC1},--\eq{MC4} and  \eq{MC5} --\eq{MC8}. 

\item[]$\mathbf{p=1,\,2,\,5}$: the algebra is given by the whole algebra \eq{MC1}--\eq{MC8}.  \end{itemize}

This last case represents the most general solution of \eq{dC}, using the algebra \eq{MC1}--\eq{MC8}. In particular, the corresponding expansion of $C$ \eq{Cparam} gives the general structure for a candidate superfivebrane Wess--Zumino term in flat background.\\
In fact, as discussed in \cite{deAzcarraga:2005jd}, the problem of reducing a minimal FDA to an ordinary algebra is mathematically equivalent to that of obtaining strictly invariant Wess--Zumino terms from the originally quasi--invariant ones. In \cite{Bergshoeff:1995hm,Sezgin:1995bh,Chryssomalakos:1999xd},  strictly invariant Wess--Zumino terms were proposed; as they include the generators $\S_{\a\b}$ they are not suitable to describe  the minimal $D=11$ FDA, for which $\mathcal{R}^{ab}=0$ .\\ 
We conclude therefore that being the $M$--Algebra  \cite{Sezgin:1996cj} the most general extension of the super translational algebra \eq{dVf}--\eq{dpf} and being \eq{MC1}--\eq{MC8} its biggest In\"on\"u--Wigner contraction allowing for commuting translations and hence for a flat background, we found the most general solution to the problem of reducing the minimal $D=11$ FDA to an algebra.\\
At this point, it is natural to consider what happens in the case of non--flat backgrounds, that is when we allow non--commuting translations, and therefore we must consider the whole $M$--Algebra. In this case one has to reduce the general $D=11$ FDA, including the contractible generators (field strengths) on the r.h.s of \eq{DV}--\eq{Dp}, \eq{dC}. Some considerations on the possibility to reduce  a non minimal FDA  to an algebra will be discussed in the next section. 


 \section{Reduction of $D=11$ FDA and the $E_{11}$ conjecture}\label{contr} 
 In order to consider non zero curvature and field strengths,  we need to reintroduce the Lorentz generators and  modify the minimal FDA  \eq{DV}--\eq{Dp}, \eq{dC} by introducing contractible generators:
\ba
&&DV^a-\frac i2\pb\g^a\p=T^a\label{T}\\
&&d\o^{ab}-\o^a_{~c}\o^{cb}=R^{ab}\label{R}\\
&&D\p=\r\label{r}\\
&&dC-\frac 12 \pb\g_{ab}\p V^aV^b=F\label{F}
\ea
The integration conditions of \eq{T}--\eq{F} are easily obtained
\ba
&&DT^a+\mathcal{R}^{ab}V_b-i\pb\g^a\r=0\label{DT}\\
&&D\mathcal{R}^{ab}=0\label{DR}\\
&&d\r+\frac14\mathcal{R}^{ab}\g_{ab}\p=0\label{Dr}\\
&&dF+ \pb\g_{ab}\p T^aV^b- \pb\g_{ab}\r V^aV^b =0\label{DF}
\ea
Usually one interprets \eq{T}--\eq{F} as a "deformation"  of the minimal FDA, where the curvature on the r.h.s. represent the fluctuations of the fields on the vacuum described by the minimal FDA, and the integration conditions \eq{DT}--\eq{DF} represent their Bianchi identities.\\   
It is clear that the group manifold $\tilde\mathcal{G}_0$, whose cotangent space $\mathcal{T}\tilde\mathcal{G}_0$  is spanned by 
\be\{V^a,\,B^a,\,B^{ab},\,B^{a_1\dots a_5},\,\p,\,\eta,\,\eta^a,\,\eta^{a_1\dots\a_4}\}\equiv\{\s^i\}\,,\label{cobasis}\ee
where the one--forms \eq{cobasis} satisfy \eq{MC1}--\eq{MC8}, is unsuitable to describe \eq{T}--\eq{DF}.\\
The standard approach \cite{cubo} is to deform the group manifold $\tilde\mathcal{G}_0$ to a ''soft group manifold'' $\tilde\mathcal{G}_0^{(soft)}$. The cotangent bundle of the soft group manifold $\mathcal{T}^*\tilde\mathcal{G}_0^{(soft)}$ is spanned by the same generators \eq{cobasis} of $\mathcal{T}^*\tilde\mathcal{G}_0$ but the left invariance condition is relaxed, that is they do not fulfill the Maurer--Cartan equations \eq{MC1}--\eq{MC8}.  Instead they satisfy
\be d\tilde\s^i-\frac 12 c^i_{\ kj}\tilde\s^j\tilde\s^k=F^i\label{rdef}\ee
where we have denoted  by $\{\tilde\s^i\}$ the soft forms the basis of $\mathcal{T}^*\tilde\mathcal{G}_0^{(soft)}$.  The presence of a curvature term on the r.h.s. has a counterpart in the appearance of  curvature terms in the minimal FDA  \eq{DV}--\eq{Dp},  \eq{dC}, thus giving \eq{T}--\eq{F}.
The curvature in \eq{rdef} can be expanded on  $\mathcal{T}^*\tilde\mathcal{G}_0^{(soft)}$  according to
\be F^i=F^i_{j_1\dots j_n}\tilde\s^{j_1}\dots\tilde\s^{j_n} \label{rpar}\ee
where the $F^i_{j_1\dots j_n}$ depend on the coordinates of $\tilde\mathcal{G}_0^{(soft)}$. The expansion \eq{rpar} fulfills the Bianchi identities \eq{DT}--\eq{DF} provided the $F^i_{j_1\dots j_n}$ satisfies some differential relations which turn out to be the equations of motion once a Lagrangian formulation is given.\\ The parametrization \eq{rpar} it is known as rheonomic parametrization and encodes the supersymmetry transformation laws. A short account is given in appendix \ref{reho}.

As pointed out in \cite{Fre:1984pc}, given \eq{T}--\eq{DF}, from the mathematical point of view it is more appropriate to consider $T^a$, $\mathcal{R}^{ab}$, $\r$, $F$ as  contractible generators which extend the minimal FDA \eq{DV}--\eq{Dp}, \eq{dC}; consequently \eq{DT}--\eq{DF} are further equations of the FDA on the same footing as  \eq{T}--\eq{F}.\\ 
Within this approach it is quite natural to wonder if the FDA \eq{T}--\eq{DF} is equivalent to an algebra obtained by further extending $\mathcal{T}\tilde\mathcal{G}_0$, which reduces to  $\mathcal{T}\tilde\mathcal{G}_0$  when the contractible generators are set to zero. This would imply that there exist a group manifold $\mathcal{G}_K \supset \tilde\mathcal{G}_0$ such that we can expand the contractible generators $F^i$ on a basis $\{\o_I\}\supset\{\s^i\}$, $I=1,\dots\, dim\,\mathcal{G}_K$,  of $\mathcal{T^*G}_K $ 
\be F^ i=f^i_{J_1\dots J_n} \o^{J_1}\dots\o^{J_n}\ee 
with constant $f^i_{J_1\dots J_n} $.

A good candidate would be therefore the whole $M$--Algebra \cite{Sezgin:1996cj}, since the next central extension of \eq{QQ+}--\eq{QZ5} will introduce e.g. the generator $\S_{\a\b}$ which are non zero in a non trivial background. Nevertheless, it is easy to show that the $M$--Algebra is certainly not sufficient to describe, for instance, the contractible generator $F$ \eq{F}.\\ 
To show this, let us preliminary observe that in the FDA \eq{T}--\eq{DF} we can set consistently to zero any of the contractible generators; therefore we can limit ourself to studying the case where only $F$ is present. This means that since the $\S_{\a\b}$ generators are related to the presence of $R^{ab}$, they should play no crucial role in the decomposition of $F$. 

In order to understand this point, let us have a closer look at the expansion of $C$ in \eq{Cparam}: if we denote by $\{Z^M\}$ the coordinates on $\tilde\mathcal{G}_0$, the one--forms \eq{cobasis} can be expressed in components as
\be V^a=V^a_M dZ^M,\quad B^{ab}=B^{ab}_MdZ^M,\dots\ee
and the space--time components of $C$ can be easily read off
\be C=dx^\m\wedge dx^\n\wedge dx^\r(\a_0 B_{\m\n\r}+\a_1B_{\m\s_1\s_2}B_\n^{~\s_2\s_3}B_{\r\s_3}^{~~\s_1}+\dots)\label{Cspace}\ee
 where \be B_{\m\n\r}\equiv B_{ab\m}V^a_\n V^b_{\r}\,.\label{Bcomp}\ee
If we include further generators, it is plausible to expect that extra terms proportional to these latter may arise in the decomposition of $C$ and especially that they will take part in the decomposition of $F$.\\
Focusing on the bosonic part of $F$ we see from \eq{F} and \eq{DF} that it has to satisfy
\be F_{|bos}=dC_{|bos}\,;\quad \quad dF_{|bos}=0 \ee
where with ''${}_{|bos}$'' we indicate the restriction to the bosonic terms.\\
It is immediate to see that every three--form $C$ one can construct using as building blocks the one--forms \eq{cobasis} satisfying \eq{MC1}--\eq{MC8} will always give $dC_{|bos}=0$, thus being unsuitable for our purposes, as we clearly cannot accept that $F$ has no bosonic space--time components.\\
It is also easy to see that the introduction of one--forms dual to the generators $\S_{\a\b}$ and $\S_{\a\b}^{abc}$,  or generators in subsequent  extensions of the $M$--Algebra would not be helpful, as we expected.\\
Indeed, if we consider a further term $\Delta C$ in the expansion of $C$ \eq{Cparam} like
\be \Delta C\propto\g^{\a\b}_a\S_{\a\b}B^{ab}V_b\ee
being 
\be d\S_{\a\b}=-\frac12V_aV_b\g_{\a\b}^{ab}+\frac12 B_{ab}V^a\g_{\a\b}^b+\frac i4 \eta_{a\g}\p^\g\g_{\a\b}^a+i\eta_{a\,\a}\p^\g\g^a_{\b\g}+i\eta_{a\,\b}\p^\g\g^a_{\a\g}\ee
we would obtain contributions to $F_{|bos}$ of the form
\be F_{|bos}\propto B_{ab}B^{bc}V^aV_c\,.\ee
If this were the case there would not exist a limit in which setting  $F$ to zero we would retrieve the algebra \eq{MC1}--\eq{MC5}. \\
In order to give appropriate contributions to $F_{|bos}$, we would need 
\be dV^{a}_{|bos}\neq0;\quad\quad dB^{a}_{|bos}\neq0;\quad\quad dB^{ab}_{|bos}\neq0;\quad\quad dB^{a_1\dots a_5}_{|bos}\neq0\ee
where the nonzero term on the r.h.s. has to contain new generators other than \eq{cobasis}, which implies that at least one generator among $P_{a}$,  $Z_{a}$ , $Z_{ab}$  and $Z_{a_1\dots a_5}$ has to arise as the commutator of two bosonic generators, which is not the case in the $M$--Algebra \cite{Sezgin:1996cj}.\\
A plausible scenario is the action of an automorphism group on the $M$--Algebra. Indeed, consider  the automorphism algebra of \eq{PP}--\eq{QQ+}: if we define a generator $Z_{\a\b}$ symmetric in the spinorial indexes
\be Z_{\a\b}= iP_a \g^a_{\a\b} +iZ_a \g^a_{\a\b} + Z_{ab} \g^{ab}_{\a\b} +iZ_ {a_1\dots a_5}\g^{a_1\dots a_5}_{\a\b} \label{gen1}\ee
we can rewrite  \eq{PP}--\eq{QQ+} in a compact form
\be \{Q_\a,Q_\b\}=Z_{\a\b}\label{QQcomp}\ee
Consider the action of a generator $R_\a^{\ \b}$ on \eq{QQcomp}, \cite{West:2004iz}:
\be [Q_\a,R_{\g}^{\ \d}]=\d_\a^\d Q_\g;\quad\quad  [Z_{\a\b},R_{\g}^{\ \d}]=\d_\a^\d Z_{\g\b} +\d_\b^\d Z_{\g\a}\label{RQRZ} \ee
which further satisfies:
\be [R_\a^{\ \b},R_\g^{\ \d}]=\d^\a_\d R_\g^{\ \b}+ \d^\b_\g R_\d^{\ \a}\label{RR}\ee
As $R_\a^{\ \b}$ does not have a definite symmetry, it can be expanded as
\be R_\a^{\ \b}=(\g^{a_1\dots a_q})_\a^{\ \b}\ R_ {a_1\dots a_q};\quad\quad q=0,\dots 10\ee 
One can see e.g. that for $q=2$ one retrieves the action of the Lorentz generators $M_{a_1a_2}$.  If we interpret  $\{P_a,\,Z_{ab},\,Z_{a_1\dots a_5}\}$ as the completely antisymmetric generators at the levels\footnote{according to the labels of \cite{Nicolai:2003fw}.}  $\ell=7$,  $8$, $9$ of $E_{11}$ we can see that the action of the generators $M_{a_1a_2}$ $R_{a_1a_2a_3}$ and $R_{a_1\dots a_6}$ at the levels $\ell=0$, $1$, $2$ respectively, is given by \cite{West:2003fc}:\ba
&&D B^{a_1a_2}_{|bos}+A^{a_1a_2a_3}V_{a_3}=0\label{caz1}\\
&&D B^{a_1\dots a_5}_{|bos}+ A^{a_1\dots a_6}V_{a_6}+ A^{[a_1a_2a_3}B^{a_4a_5]}=0\label{caz2}\\
&&D A^{a_1a_2a_3}_{|bos}=0\label{caz3}\\
&&D A^{a_1\dots a_6}_{|bos}+ A^{[a_1a_2a_3}A^{a_4a_5a_6]}=0\label{caz4}
\ea
where we have defined the dual one-forms according to
\be A^{a_1\dots a_3}R_{b_1\dots b_3}=\d^{a_1\dots a_3}_{b_1\dots b_3};\quad\quad A^{a_1\dots a_6}R_{b_1\dots b_6}=\d^{a_1\dots a_6}_{b_1\dots b_6}\label{dualblu}\ee
and $D$ is defined in \eq{coder}.

In this case the bosonic components of $F$, \eq{F}, can be obtained by differentiating \eq{Cparam} and using \eq{caz1} and \eq{caz2}. The space--time components are read off as before:
\be F_{|bos}=dx^\m dx^\n dx^\r dx^\t (-\a_0A_{\m\n\r\t}-3\a_1 A_{\m\n\s_1\s_2}B_\r^{~\s_2\s_3}B_{\t\s_3}^{~~\s_1}+\dots)\label{Fspace}\ee
with 
\be A_{\m\n\r\t}\equiv A_{abc\m}V^a_\n V^b_\r V^c_\t\,.\label{Fcomp}\ee 
The bosonic Bianchi identity $dF_{|bos}=0$ is a consequence of the closure of the algebra \eq{caz1}--\eq{caz4}. 

The possibility to use \eq{caz1}--\eq{caz4} to describe the bosonic $D=11$ FDA in the presence of the contractible generator $F$, is intriguing as it suggests that non trivial backgrounds of $D=11$ supergravity enjoy (low level) $E_{11}$ symmetry and that this is encoded in the $D=11$ FDA \eq{T}--\eq{DF}.

Unlike the non linear realizations where the generators $R_{a_1a_2a_3}$ and $R_{a_1\dots a_6}$ are associated to the three--form $C$ and its dual six--form $\tilde C$ respectively, we interpret  $R_{a_1a_2a_3}$ and $R_{a_1\dots a_6}$ and  their dual one--forms $A^{a_1a_2a_3}$ and $A^{a_1\dots a_6}$  \eq{dualblu}  as elements of $\mathcal{TG}_K$ and $\mathcal{T^*G}_K$ respectively, where $\mathcal{G}_K$ is a  group manifold on which $D=11$ supergravity is formulated. The one--forms $A^{a_1a_2a_3}$ and $A^{a_1\dots a_6}$ are not seen as physical fields; instead they are part of the composite structure of $F$, just like the one--forms \eq{cobasis} of the composite structure of $C$ \cite{Bandos:2004ym}.  

\section{Conclusions and outlook}\label{specul}

In this paper we reconsidered the problem of reducing the Free Differential Algebra of $D=11$ supergravity to an ordinary  algebra. We addressed separately the case of the minimal  and the general FDA, where the contractible generators represent the field strengths. 

For the minimal FDA, instead of looking for the simplest solution as it was done so far, we tried to find the most general one. For this reason we considered as candidate algebra the $M$--Algebra, which is the biggest extension of the $D=11$ super translational algebra. In this respect we considered its  biggest In\"on\"u--Wigner contraction  admitting commuting translations \eq{QQ+}--\eq{QZ5} and showed the equivalence with the minimal $D=11$ FDA.\\
It is interesting to observe that the $D=11$ FDA already encodes information on non perturbative states like the M2 and M5 branes, as one can see from equation \eq{QQ+}.  It is in fact well known \cite{deAzcarraga:1989gm} that the presence of supersymmetric extended objects modifies the super Poincar\'e algebra by introducing topological charges, e.g. \eq{QQ+}. 

For the general FDA we limited ourself to take in consideration the contractible generator $F$ and to study the problem at the bosonic level. We found that a convincing scenario is to consider the action of the automorphism group  proposed in \cite{West:2004iz}  and we showed that  in a flat background the lowest levels of $E_{11}$ can describe the reduction of the bosonic $D=11$ FDA to an algebra.

This partial result suggests that the $E_{11}$ symmetry, which acts as well on perturbative states,  is already encoded in the $D=11$ FDA and is made explicit when the theory is formulated on a suitable group manifold $\mathcal{G}_K$, using as vector potentials a basis of  one--forms on the cotangent space $\mathcal{T^*G}_K$.   

To complete the picture, there is a certain number of issues that need to be addressed. The first one would obviously be to go through  the reduction of the FDA in the presence of the sole contractible generator $F$ considering also the fermionic sector; that is, considering the action of the automorphism group on the whole  $\tilde\mathcal{G}_0$. Afterwards,  to address the problem in the presence of all the contractible generators. 

Another important point is that,  contrary to the rheonomic parametrizations on the soft group manifold, the formulation on the enlarged group $\mathcal{G}_K$  does not enforce the equations of motion. If we want to recover this piece of information, we need a democratic formulation of the $D=11$ FDA, where for each field, included for the gravitational d.o.f., the corresponding dual is introduced. In case the democratic $D=11$ FDA can be reduced to an algebra, this would encode all the dynamics of $D=11$ supergravity. This last point would be of particular relevance as it would make unnecessary the existence of an action.\\
We hope to report soon on these issues  \cite{prep} .

\begin{center}
\bf{Acknowledgments}
\end{center}     
The author would like to thank Laura Andrianopoli, Jos\'e A. de Azc\'arraga, Sophie De Buyl,  Paul Cook, Riccardo D'Auria,  Olaf Hohm, Carlo Iazeolla, Patrick Meessen,Tom\'as Ort\'\i n, Henning Samtleben, Mario Trigiante  for their valuable help and criticism.

This work has been supported in part by the Spanish Ministry of Science 
and Education grant BFM2003-01090, the Comunidad de Madrid grant HEPHACOS 
P-ESP-00346 and by the EU Research Training Network {\em Constituents, 
Fundamental Forces and Symmetries of the Universe} MRTN-CT-2004-005104

Part of this work was carried out at the II. Institut f\"ur Theoretische Physik of the University of Hamburg,  supported by 
GIF --The German-Israeli-Foundation under Contract No. I-787-100.14/2003, DFG -- The German Science Foundation, the European RTN Programs MRTN-CT-2004-005104 and MRTN-CT-2004-503369 and the DAAD -- the German Academic Exchange Service. 


\appendix

\section{}\label{FDA}
In this appendix we present the Fierz identities that have been used in the paper.\\
We indicate with $\Xi_{a_1\dots\a_5}^{(4224)}$, $\Xi_{a_1a_2}^{(1408)}$, $\Xi_{a}^{(320)}$, $\Xi^{(32)}$  the irreducible ${\rm SO}(1,10)$ fermionic representations $(\frac32)^5$, $(\frac32)^2(\frac12)^3$, $(\frac32)(\frac12)^4$, $(\frac12)^5$ respectively \cite{D'Auria:1982nx}
\be\g^{a_1}\Xi_{a_1\dots\a_5}^{(4224)}=\g^{a_1}\Xi_{a_1a_2}^{(1408)}=\g^{a}\Xi_{a}^{(320)}=0\ee

\ba
\p\pb\g_a\p\!\!&=&\!\!\Xi_a^{(320)}+\frac1{11}\g_a\Xi^{(32)}\\
\p\pb\g_{ab}\p\!\!&=&\!\!\Xi_{ab}^{(1408)}-\frac29\g_{[a}\Xi_{b]}^{(320)}+\frac1{11}\g_{ab}\Xi^{(32)}\\
\p\pb\g_{a_1\dots a_5}\p\!\!&=&\!\!\!\Xi_{a_1\dots\a_5}^{(4224)}\!+\!2\g_{[a_1a_2a_3}\Xi_{a_4a_5]}^{(1408)}\!+\!\frac59\g_{[a_1\!\dots a_4}\Xi_{a_5]}^{(320)}\!-\!\frac1{77}\g_{a_1\dots a_5}\Xi^{(32)}
\ea
\ba
\g^a\p\pb\g_a\p\!\!&=&\!\!\Xi^{(32)}\\
\g^{ab}\p\pb\g_{ab}\p\!\!&=&\!\!-10\,\Xi^{(32)}\\
\g^{a_1\dots a_5}\p\pb\g_{a_1\dots a_5}\p\!\!&=&\!\!-720\,\Xi^{(32)}\\
\g_{ab}\p\pb\g^b\p\!\!&=&\!\!-\Xi_a^{(320)}+\frac{10}{11}\g_a\Xi^{(32)}\\
\g^b\p\pb\g_{ab}\p\!\!&=&\!\!\Xi_a^{(320)}-\frac{10}{11}\g_a\Xi^{(32)}\\
\g^{a_5}\p\pb\g_{a_1\dots a_5}\p\!\!&=&\!\!6\g_{[a_1a_2}\Xi_{a_3a_4]}^{(1408)}\!+\!\frac{24}{9}\g_{[a_1a_2a_3}\Xi_{a_4]}^{(320)}\!-\!\frac1{11}\g_{a_1\dots a_4}\Xi^{(32)}\\
\g_{a_1\dots a_5}\p\pb\g^{a_5}\p\!\!&=&\!\!-4\g_{[a_1a_2 a_3}\Xi_{a_4]}^{(320)}+\frac7{11}\g_{a_1\dots a_4}\Xi^{(32)}\\
\g_{[a_1a_2}\p\pb\g_{a_3a_4]}\p\!\!&=&\!\!\g_{[a_1a_2}\Xi_{a_3a_4]}^{(1408)}-\frac{2}{9}\g_{[a_1a_2a_3}\Xi_{a_4]}^{(320)}+\frac1{11}\g_{a_1\dots a_4}\Xi^{(32)}
\ea

\ba
&&\g^a\p\pb\g_a\p=-\frac1{10}\g^{ab}\p\pb\g_{ab}\p=-\frac1{720}\g^{a_1\dots a_5}\p\pb\g_{a_1\dots a_5}\p\label{fierz1}\\
&&\g_{ab}\p\pb\g^b\p+\g^b\p\pb\g_{ab}\p=0\label{fierz2}\\
&&\g^{a_5}\p\pb\g_{a_1\dots a_5}\p+\g_{a_1\dots a_5}\p\pb\g^{a_5}\p-6\g_{[a_1a_2}\p\pb\g_{a_3a_4]}\p=0\label{fierz3}
\ea

\ba&& \pb\g_a\p\pb\g^{ab}\p=0\label{4lin1}
\ea

\section{}\label{sol}

In this appendix we present the system of equation obtained inserting the ansatz \eq{Cparam} into equation \eq{dC} and its solutions for different values of $p$.

\subsection*{The system:}

\ba
&&\frac{{{\alpha }_0}}{2} - 3024{{\pi }_1}{{\beta }_5} + 
    9{{\beta }_1}{{\gamma }_1} + {\hat\b_1}{{\delta }_0} = \frac12\label{ppVV}\\
&&\frac{{{\alpha }_6}}{2} - 3024{{\pi }_1}{{\beta }_9} + 9{{\beta }_8}{{\gamma }_1} + {\hat\b_4}{{\delta }_0} + {\hat\b_1}{{\delta }_3} = 0\label{ppVZ}\\
&&\frac12{\alpha }_7 + {\hat\b_4}{{\delta }_3} = 0\label{ppZZ}\\
&& -i\alpha _0- \frac{ i }{2}{{\alpha }_6} - {{\beta }_2}{{\gamma }_1} - 18{{\beta }_7}{{\gamma }_1} + {{\beta }_1}{{\gamma }_2} - 2 i {\hat\b_2}{{\delta }_0} - 2 i  {\hat\b_1}{{\delta }_1} = 0\\
&&\frac{i}{2}{{\alpha }_6} +  i  {{\alpha }_7} - {{\beta }_8}{{\gamma }_2} + 2 i {\hat\b_4}{{\delta }_1} +  2 i  {\hat\b_2}{{\delta }_3} = 0\\
&& -\frac12\hat\b_1 - 5{\hat\b_2} + 360{\hat\b_3} - \frac12\hat\b_4 = 0\\
&&  -120{{\alpha }_3} - {{\pi }_1}{{\beta }_3} - {{\pi }_2}{{\beta }_5} - {{\beta }_6}{{\gamma}_1} + {\hat\b_3}{{\delta }_0} + {\hat\b_1}{{\delta }_2} = 0\\
&&  -120{{\alpha }_5} - {{\pi }_2}{{\beta }_9} + {\hat\b_4}{{\delta }_2} + {\hat\b_3}{{\delta }_3} = 0\\
&&  \frac{3}{2}{\alpha }_1 + \frac{16}{3}\pi _5\beta _4 + {{\beta }_2}{{\gamma }_2} + 2{{\beta }_7}{{\gamma }_2} + 4{\hat\b_2}{{\delta }_1} = 0\\
 && \frac12{\alpha }_3 + \frac12\alpha _5 - {\hat\b_3}{{\delta }_2} = 0\\
 && \frac12{\alpha }_2 + {{\pi }_2}{{\beta }_3} - 600{\hat\b_3}{{\delta }_2} = 0\\
&& \frac12{\alpha }_4 + \frac53{\hat\b_3}{{\delta }_2} = 0\\
 &&  i  {{\alpha }_2} - {{\beta }_6}{{\gamma }_2} - 10 i  {\hat\b_3}{{\delta }_1} -  10 i  {\hat\b_2}{{\delta }_2} = 0\\
 && i  {{\beta }_3} -  i  \left( {{\beta }_5} + {{\beta }_9} \right)  = 0\\
 &&  i  {{\beta }_3} + \frac16{\beta }_4 = 0\\
&&  -{{\beta }_2} - 336 i {{\beta }_6} + 2{{\beta }_7} - i  \left( {{\beta }_1} + {{\beta }_8} \right)  = 0\\
&&  10{{\beta }_2} + 720 i {{\beta }_6} + 90{{\beta }_7} + 10 i \left( {{\beta }_1} + {{\beta }_8} \right)  = 0
\label{sistema}
\ea

\subsection*{The solutions:}

$\mathbf{p=2}$: the decomposition of $C$ is given by \eq{Cparam} with
\be \a_0=\frac{1}{10};\quad\a_1=-\frac{1}{30};\quad\b_1=-\frac1{10};\quad\b_2=\frac i{10}\ee
and zero otherwise.
\vskip0.7cm
\noindent$\mathbf{p=1,\,2}$: the decomposition of $C$ is given by \eq{Cparam} with
\ba
&&{\alpha }_0= \frac 1{50}(5 - 50\,\hat\b_1 + 54\,\lambda \,{\hat\b_1} - 540\,{\hat\b_2} + 540\,\lambda \,{\hat\b_2})\nn\\
&&{\alpha }_1= \frac1 {30}(-1 - 8\,\lambda \,{\hat\b_2})\nn\\
&& {\alpha }_6= 
    -\frac 2{25}\,\left( -25\,{\hat\b_1} + 26\,\lambda \,{\hat\b_1} - 260\,{\hat\b_2} + 270\,\lambda \,{\hat\b_2} \right) \nn\\
&& \alpha_7= -{\hat\b_1} + \lambda \,{\hat\b_1} - 10\,{\hat\b_2} + 10\,\lambda \,{\hat\b_2}\nn\\
 && {\beta }_1= \frac1{50} \left(-5 + 6\,\lambda \,{\hat\b_1} - 60\,{\hat\b_2} + 60\,\lambda \,{\hat\b_2}\right)\nn\\ 
&&{\beta }_2= \frac{ i}{10}\,\left( 1 + 12\,\lambda \,{\hat\b_2} \right) \nn\\
 && {\beta }_8= -\frac{3}{25}\,\left( \lambda \,{\hat\b_1} - 10\,{\hat\b_2} + 20\,\lambda \,{\hat\b_2} \right) \nn\\
&&\hat\b_4= -{\hat\b_1} - 10\,{\hat\b_2}
 \ea
and arbitrary $\l$,  $\hat\b_1$, $\hat\b_2$,  zero otherwise.

\vskip0.7cm

\noindent$\mathbf{p=2,\,5}$: the decomposition of $C$ is given by \eq{Cparam} with

\ba
&&\a_0=\frac{1}{10}-9i\b_7;\quad\a_1=-\frac{1}{30}-\frac{8}3 i\b_7;\quad\a_2=-\frac i{48}\b_7;\nn\\
&&\b_1=-\frac1{10}+6i\b_7;\quad\b_2=\frac i{10}-6\b_7;\quad\b_3=-\frac i{48}\b_7;\nn\\
&&\b_4=-\frac18\b_7;\quad\b_5=-\frac i{48}\b_7;\quad\b_6=-\frac i{24}\b_7
\ea
with arbitrary $\b_7$, zero otherwise.

\vskip0.7cm

\noindent$\mathbf{p=1,\,2,\,5}$: the decomposition of $C$ is given by \eq{Cparam} with

\ba
{\alpha }_0=&&\!\!\!\!\!\!
    \frac1{20}\left[-1+\left( - 20 + 24\lambda + 20\tau \right){\hat\b_1}+\left( - 240 + 228\lambda + 230\tau \right){\hat\b_2} \right.+\nn\\
&&\quad+\left.\left( 14400 - 16560\lambda - 16200\tau \right){\hat\b_3} - 30{\beta }_1 - 10{\beta }_8\right]\nn\\
{\alpha }_1=&&\!\!\!\!\!\!
    \frac1{90}\left[-7+\left( - 72\lambda + 30\tau\right)\hat\b_2 - \left(2160\lambda+ 7800\tau \right){\hat\b_3} - 40{\beta }_1 - 40{\beta }_8\right]\nn\\
{\alpha }_2=&&\!\!\!\!\!\! \frac1{4\cdot 6!} \left[1 +\left( 12\lambda + 10\tau\right)\hat\b_2+\left( - 720\lambda + 1800\tau\right) {\hat\b_3} + 10{\beta }_1 + 
        10{\beta }_8\right]\nn\\
{\alpha }_3=&&\!\!\!\!\!\!
    \frac2{10!}\left[ \left(-3\lambda + 8\tau\right)\hat\b_1 +\left( 30- 60\lambda + 50\tau\right) {\hat\b_2} +\right.\nn\\
&&\quad\left.+ \left(5760 -  3600\lambda - 9000\tau\right) {\hat\b_3} - 25{{\beta }_8}\right]\nn\\
 {\alpha }_4=&&\!\!\!\!\!\! -\frac1 {432}\tau {\hat\b_3}\nn\\
{\alpha }_5=&&\!\!\!\!\!\!
    \frac2{10!}\left[ \left(3\lambda - 8\tau\right) {\hat\b_1}+\left( - 30 + 60\lambda - 50\tau\right) {\hat\b_2}+\right.\nn\\
&&\quad\left.+\left( - 5760 +  3600\lambda + 11520\tau\right) {\hat\b_3} + 25{{\beta }_8}\right]\nn\\
{\alpha }_6=&&\!\!\!\!\!\!
    \frac15\left[ \left(10 - 11\lambda - 10\tau\right) {\hat\b_1} +\left( 110 - 120\lambda - 110\tau\right) {\hat\b_2}+\right.\nn\\
&&\quad\left.+\left( - 7200 + 7920\lambda + 7200\tau\right) {\hat\b_3} - 5{{\beta }_8}\right]\nn\\
 {\alpha }_7=&&\!\!\!\!\!\!\left[\left( -1 + \lambda + \tau\right) {\hat\b_1}+\left( - 10 + 10\lambda + 10\tau\right) {\hat\b_2} +\left( 720 -  720\lambda - 720\tau\right) {\hat\b_3}\right]\nn\\
{\beta }_2=&&\!\!\!\!\!\!
    \frac{i}{5}\left[1+\left( 12\lambda - 10\tau\right) {\hat\b_2} + \left(720\lambda  + 1800\tau\right) {\hat\b_3} + 5{{\beta }_1} + 5{{\beta }_8} \right] \nn\\
{\beta }_3=&&\!\!\!\!\!\!
    \frac{1}{4\cdot 6!}\left[-1+\left( - 12\lambda - 10\tau\right) {\hat\b_2} + \left(720\lambda + 600\tau\right) {\hat\b_3} - 10{{\beta }_1} - 10{{\beta }_8}\right]\nn\\
{\beta }_4=&&\!\!\!\!\!\!
    \frac{i}{480}\left[ 1 +\left( 12\lambda + 10\tau\right) {\hat\b_2} +\left(- 720\lambda - 600\tau\right) {\hat\b_3} + 10{{\beta }_1} + 10{{\beta }_8} \right]\nn\\
{\beta }_5=&&\!\!\!\!\!\!
    \frac{6}{9!}\left[-21 +\left( 24\lambda + 20\tau\right) {\hat\b_1}+\left( - 240 + 228\lambda + 230\tau\right) {\hat\b_2}\right.+\nn\\
&&\left. +\left( 14400 - 16560\lambda - 16200\tau\right) {\hat\b_3} - 210{{\beta }_1} - 10{{\beta }_8}\right]\nn\\
\beta_6=&&\!\!\!\!\!\!
\frac{1}{2\cdot 6!}\left[-1+\left(- 12\lambda + 10\tau\right)\hat\b_2+\left(- 720\lambda - 1800\tau\right) \hat\b_3 - 10\beta _1 - 10\beta _8\right]\nn\\ 
\beta_7=&&\!\!\!\!\!\! -\frac{i}{60}\left[1+\left(12\lambda - 10\tau\right) {\hat\b_2} +\left( 720\lambda + 1800\tau\right) {\hat\b_3} + 10{{\beta }_1} + 10{{\beta }_8}\right]\nn\\
{\beta }_9= &&\!\!\!\!\!\!
 \frac{1}{3\cdot 7!}\left[\left(-6\lambda - 5\tau\right) {\hat\b_1} +\left( 60 - 120\lambda - 110\tau\right) {\hat\b_2}+\right.\nn\\
&&\quad\left.+\left( - 3600 + 7920\lambda + 7200\tau\right) {\hat\b_3} - 50{{\beta }_8}\right]\nn\\
 \hat\b_4=&&\!\!\!\!\!\! -{\hat\b_1} - 10{\hat\b_2} + 720{\hat\b_3}
\label{solution}
\ea
With arbitrary $\l$, $\t$, $\b_1$, $\b_8$, $\hat\b_1$, $\hat\b_2$, $\hat\b_3$.

\section{} \label{reho}
For sake of completeness we present the rheonomic parametrization of the $D=11$ supercurvatures \cite{D'Auria:1982nx}.
\ba
T^a\!\!\!&=&\!\!\!0\label{rh1}\\
R^{ab}\!\!\!&=&\!\!\!R^{ab}_{\ \ cd}V^c V^d+i(2\bar\r_{c[a}\g_{b]}-\bar\r_{ab}\g_c)\p V^c+F^{abcd}\pb\g^{cd}\p+\nn\\
&&+\frac 1{24}F_{cdef}\pb\g^{abcdef}\p\\
\r\!\!\!&=&\!\!\!\r_{ab}V^a V^b+\frac i3(F_{abcd}\g^{bcd}-\frac18F_{bcde}\g_a^{\ bcde})\p V^a\\
F\!\!\!&=&\!\!\! F_{abcd}V^a V^bV^c V^d\label{rh4}
\ea

In order \eq{rh1}--\eq{rh4} to satisfy the Bianchi identities \eq{DT}--\eq{DF},  $R^{ab}_{\ \ cd}$, $\r_{ab}$, $F_{abcd}$  has to satisfy the propagation equations
\ba
&&R^{ac}_{\ \ bc}-\frac12 \d^a_b R-3 F^a_{\ cde}F_{bcde}+\frac 38 \d^a_b F^{cdef}F_{cdef}=0\\
&& \g^{abc}\r_{bc}=0\\
&&\partial_m F^{mabc}-\frac 1{2\cdot 4!\cdot 7!} \epsilon^{abcdefghijkl} F_{efgh} F_{ijkl}=0
\ea

The determination of the supersymmetry transformation laws from the rheonomic parametrizations is obtained considering the Lie derivative along the tangent vector
\be \epsilon=\ve\vec D\ee
where $\vec D$ is dual to the gravitino one--form $\p$. Denoting by $\m^I$ the $p$--form potentials and with $F^I$ the $p+1$--forms field strengths, one has:
\be\ell\m^I=(i_\e\, d+d\, i_\e)\m^I \equiv (D\e)^I+i_\e F^I\ee
where $D$ is the covariant derivative \eq{coder} and $i_\e$ is the contraction operator along the vector $\e$. 

\appendixend


\begin{thebibliography}{99}

\bibitem{Cremmer:1978km}
  E.~Cremmer, B.~Julia and J.~Scherk,
  Phys.\ Lett.\ B {\bf 76} (1978) 409.
  
  
\bibitem{Cremmer:1978ds}
  E.~Cremmer and B.~Julia,
  ``The N=8 Supergravity Theory. 1. The Lagrangian,''
  Phys.\ Lett.\ B {\bf 80} (1978) 48.
  
\bibitem{Julia:1981wc}
  B.~Julia,
  ``Infinite Lie Algebras In Physics,''
LPTENS-81-14
{\it Invited talk given at Johns Hopkins Workshop on Current Problems in Particle Theory, Baltimore, Md., May 25-27, 1981}



\bibitem{Cremmer:1998px}
  E.~Cremmer, B.~Julia, H.~Lu and C.~N.~Pope,
  ``Dualisation of dualities. II: Twisted self-duality of doubled fields  and superdualities,''
  Nucl.\ Phys.\ B {\bf 535} (1998) 242
  [arXiv:hep-th/9806106].


\bibitem{West:2000ga}
  P.~C.~West,
  ``Hidden superconformal symmetry in M theory,''
  JHEP {\bf 0008} (2000) 007
  [arXiv:hep-th/0005270].

\bibitem{West:2001as}
  P.~C.~West,
  ``E(11) and M theory,''
  Class.\ Quant.\ Grav.\  {\bf 18} (2001) 4443
  [arXiv:hep-th/0104081].


\bibitem{Damour:2005zs}
  T.~Damour, A.~Kleinschmidt and H.~Nicolai,
  ``Hidden symmetries and the fermionic sector of eleven-dimensional supergravity,''
  Phys.\ Lett.\ B {\bf 634} (2006) 319
  [arXiv:hep-th/0512163].


\bibitem{deBuyl:2005mt}
  S.~de Buyl, M.~Henneaux and L.~Paulot,
  ``Extended E(8) invariance of 11-dimensional supergravity,''
  JHEP {\bf 0602} (2006) 056
  [arXiv:hep-th/0512292].

\bibitem{Damour:2006xu}
  T.~Damour, A.~Kleinschmidt and H.~Nicolai,
  ``K(E(10)), supergravity and fermions,''
  JHEP {\bf 0608} (2006) 046
  [arXiv:hep-th/0606105].


\bibitem{Damour:2000wm}
  T.~Damour and M.~Henneaux,
  ``Chaos in superstring cosmology,''
  Phys.\ Rev.\ Lett.\  {\bf 85} (2000) 920
  [arXiv:hep-th/0003139].

\bibitem{Damour:2002et}
  T.~Damour, M.~Henneaux and H.~Nicolai,
  ``Cosmological billiards,''
  Class.\ Quant.\ Grav.\  {\bf 20} (2003) R145
  [arXiv:hep-th/0212256].

\bibitem{Bergshoeff:2006qw}
  E.~A.~Bergshoeff, M.~de Roo, S.~F.~Kerstan, T.~Ortin and F.~Riccioni,
  ``IIA ten-forms and the gauge algebras of maximal supergravity theories,''
  JHEP {\bf 0607} (2006) 018
  [arXiv:hep-th/0602280].






\bibitem{D'Auria:1982nx}
  R.~D'Auria and P.~Fre,
  ``Geometric Supergravity In D = 11 And Its Hidden Supergroup,''
  Nucl.\ Phys.\ B {\bf 201} (1982) 101
  [Erratum-ibid.\ B {\bf 206} (1982) 496].



\bibitem{sullivan}
D. Sullivan, Infinitesimal computations in topology, reprints from IHES, Bures-sur-Yvette, France. 

\bibitem{cubo}
The reference text for the rehonomic approach to the construction of supergravity is:\\ 
R.D'Auria, L.Castellani and P.Fre`,``Supergravity and
Superstrings: A Geometric Perspective'',Vol.1,\,2, World Scientific
1991.
There are several good lectures on the topic, e.g.:
L.~Castellani,
  ``Group Geometric Methods In Supergravity And Superstring Theories,''
  Int.\ J.\ Mod.\ Phys.\ A {\bf 7} (1992) 1583.




\bibitem{Sezgin:1996cj}
  E.~Sezgin,
  ``The M-algebra,''
  Phys.\ Lett.\ B {\bf 392} (1997) 323
  [arXiv:hep-th/9609086].

\bibitem{Fre:1984pc}
  P.~Fre,
  ``Comments On The Six Index Photon In D = 11 Supergravity And The Gauging Of Free Differential Algebras,''
   Class.\ Quant.\ Grav.\  {\bf 1} (1984) L81.


\bibitem{Barwald:1999is}
  O.~Barwald and P.~C.~West,
  ``Brane rotating symmetries and the fivebrane equations of motion,''
  Phys.\ Lett.\ B {\bf 476} (2000) 157
  [arXiv:hep-th/9912226].

\bibitem{West:2004iz}
  P.~West,
  ``Brane dynamics, central charges and E(11),''
  arXiv:hep-th/0412336.






\bibitem{Bandos:2004xw}
  I.~A.~Bandos, J.~A.~de Azcarraga, J.~M.~Izquierdo, M.~Picon and O.~Varela,
  ``On the underlying gauge group structure of D = 11 supergravity,''
  Phys.\ Lett.\ B {\bf 596} (2004) 145
  [arXiv:hep-th/0406020].

  
\bibitem{Bandos:2004ym}
  I.~A.~Bandos, J.~A.~de Azcarraga, M.~Picon and O.~Varela,
  ``On the formulation of D = 11 supergravity and the composite nature of  its three-form field,''
  Annals Phys.\  {\bf 317}, 238 (2005)
  [arXiv:hep-th/0409100].
  
  

  
  
\bibitem{Castellani:1995gz}
  L.~Castellani and A.~Perotto,
  ``Free Differential Algebras: Their Use in Field Theory and Dual Formulation,''
  Lett.\ Math.\ Phys.\  {\bf 38} (1996) 321
  [arXiv:hep-th/9509031].
  
\bibitem{Castellani:2005vt}
  L.~Castellani,  ``Lie derivatives along antisymmetric tensors, and the M-theory superalgebra,''
  arXiv:hep-th/0508213.

  
\bibitem{Castellani:2006jg}
  L.~Castellani,
  ``Extended Lie derivatives and a new formulation of D = 11 supergravity,''
  arXiv:hep-th/0604213.
 
 
 




\bibitem{deAzcarraga:2005jd}
  J.~A.~de Azcarraga,
   ``Superbranes, D = 11 CJS supergravity and enlarged superspace  coordinates 
  fields correspondence,''
  AIP Conf.\ Proc.\  {\bf 767} (2005) 243
  [arXiv:hep-th/0501198].


 
\bibitem{Bergshoeff:1995hm}
  E.~Bergshoeff and E.~Sezgin,
  ``Superp-Brane theories and new space-time superalgebras,''
  Phys.\ Lett.\ B {\bf 354} (1995) 256
  [arXiv:hep-th/9504140].
   
\bibitem{Sezgin:1995bh}
  E.~Sezgin,
  ``Super p-Form Charges and a Reformulation of the Supermembrane Action in Eleven Dimensions,''
  arXiv:hep-th/9512082.
  
\bibitem{Chryssomalakos:1999xd}
  C.~Chryssomalakos, J.~A.~de Azcarraga, J.~M.~Izquierdo and J.~C.~Perez Bueno,
  ``The geometry of branes and extended superspaces,''
  Nucl.\ Phys.\ B {\bf 567} (2000) 293
  [arXiv:hep-th/9904137].
  



\bibitem{Nicolai:2003fw}
  H.~Nicolai and T.~Fischbacher,
  ``Low level representations for E(10) and E(11),''
  arXiv:hep-th/0301017.

\bibitem{West:2003fc}
  P.~West,
 ``E(11), SL(32) and central charges,''
  Phys.\ Lett.\ B {\bf 575} (2003) 333
  [arXiv:hep-th/0307098].


\bibitem{deAzcarraga:1989gm}
  J.~A.~de Azcarraga, J.~P.~Gauntlett, J.~M.~Izquierdo and P.~K.~Townsend,
  `` Topological Extensions Of The Supersymmetry Algebra For Extended Objects''
  Phys.\ Rev.\ Lett.\  {\bf 63} (1989) 2443.


  
  \bibitem{prep} Silvia Vaul\`a, in preparation.


\end{thebibliography}
\end{document}